\documentclass[preprint,aps,nofootinbib,tightenlines]{revtex4}

\usepackage{amsmath,amssymb,amsbsy,mathbbol}
\usepackage{epsfig}
\usepackage{bm}

\def\cL{{\mathcal L}}

\begin{document}

\preprint{\vbox{
\hbox{CALT-68-2589}
}}
\vphantom{}

\title{Regulator Dependence of the Proposed UV Completion of the Ghost Condensate}
\author{Donal O'Connell}
\affiliation{
Department of Physics, California Institute of Technology,
Pasadena, CA 91125, USA.}

\vphantom{}
\vskip 0.5cm
\begin{abstract} 
\vskip 0.5cm
\noindent 
Recently, it was shown that a renormalizable theory of heavy fermions
coupled to a light complex boson could generate an effective action for
the boson with the properties required to violate Lorentz invariance
spontaneously through the mechanism of ghost condensation. However,
there was some doubt about whether this result depended on the choice of
regulator. In this work, we adopt a non-perturbative, unitary lattice
regulator and show that with this regulator the theory does not have
the properties necessary to form a ghost condensate. Consequently,
the statement that the theory is a UV completion of the Higgs phase of
gravity is regulator dependent.

\end{abstract}

\maketitle

\section{Introduction}

\begin{figure}[b]
\includegraphics{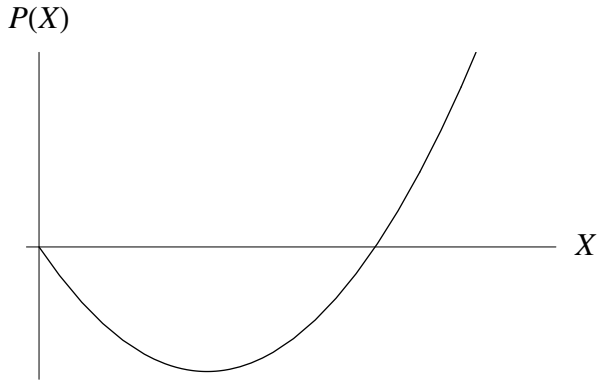}
\caption{One possible form for $P(X)$.}
\label{fig:p1}
\end{figure}

\begin{figure}[t]
\includegraphics{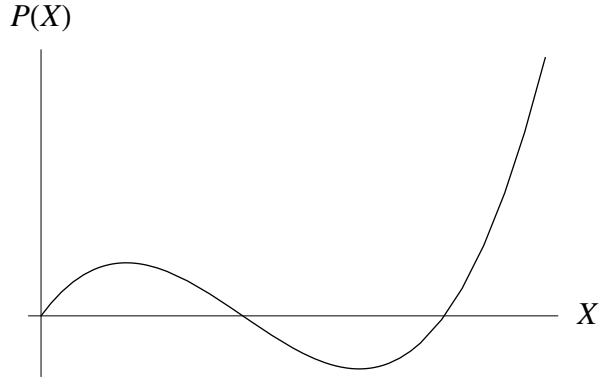}
\caption{Another possible form for $P(X)$, with no ghost at the origin.}
\label{fig:p2}
\end{figure}

The ghost condensate proposal of~\cite{Arkani-Hamed:2003uy} has
received considerable attention
recently~\cite{Graesser:2005ar,Arkani-Hamed:2003uz,Dubovsky:2004qe,Peloso:2004ut,Holdom:2004yx,Frolov:2004vm,
Arkani-Hamed:2004ar,Krotov:2004if,Mukohyama:2005rw,Arkani-Hamed:2005gu,Anisimov:2004sp}.
The condensate is a mechanism for modifying gravity in the infrared.
The starting point of the model is a scalar field, $\phi$, with a
shift symmetry
\begin{equation}
\phi \rightarrow \phi + \alpha
\end{equation}
such that the effective action for the scalar is of the form $\cL =
P(X)$, where $X = \partial_\mu \phi \partial^\mu \phi$ (we ignore terms
such as $(\partial^2 \phi)^2$ as they will not be important in our
discussion.) Moreover, we assume that $\phi$ is a ghost, so that $P(X)$
is of the form shown in Figure~\ref{fig:p1}. The origin, $\phi = 0$, is an
unstable field configuration in this scenario. The ghost then condenses
so that $(\partial \phi)^2$ has a value near the minimum of $P$. It is
also possible that there is no ghost at the origin but a non-trivial
minimum elsewhere, as shown in Figure~\ref{fig:p2}; in such a theory
there would still be a ghost condensate near the minimum of $P$. This
class of theories is of considerable phenomenological interest because
a ghost condensate has equation of state $w = -1$ and could therefore
be relevant for explaining the observed small but non-zero cosmological
constant~\cite{Arkani-Hamed:2003uy}.

It is also of interest, however, to understand how the effective action
$\cL = P(X)$ could arise as a low energy effective theory of some more
familiar UV quantum field theory~\cite{Krotov:2004if}. Since the scalar
field must have a shift symmetry, it is natural to seek a completion
in which $\phi$ is the Goldstone boson of a spontaneously broken $U(1)$
symmetry.
It was shown in~\cite{Graesser:2005ar} that it is impossible,
classically, to generate a ghostly low energy effective action for such a
Goldstone boson from a high energy theory with standard kinetic terms.
However, the authors went on to find a theory in which a quantum
correction could change the sign of the kinetic term of the Goldstone boson.
In that proposal, all fields start out with standard kinetic
terms. However, interactions between $\phi$ and certain heavy fermions
correct the kinetic term of $\phi$. It was found that under certain
assumptions, these corrections could produce an effective Lagrangian
for $\phi$ of the form shown in Figure~\ref{fig:p1} at scales much
smaller than the fermion mass $m$. We do not expect to find an effective
Lagrangian of the form shown in Figure~\ref{fig:p2} because the higher
order terms in the expansion of $P(X)$ are suppressed by powers of
the cutoff.

The model described in~\cite{Graesser:2005ar} has some shortcomings. The
high energy theory has a Landau pole. Moreover, in dimensional
regularization it was found that to change the sign of the bosonic
kinetic term, the mass of the fermions has to be close to the Landau
pole. This circumstance may cause some concern that the calculation
could be regulator dependent. To alleviate these concerns, the
authors demonstrated that their conclusion holds in a large class
of momentum-dependent regulators, provided that the fermion masses
were taken to be of order of the regulator. These regulators, however,
violate unitarity, so again it is not clear to what extent the sign of
the kinetic term is a well defined quantity.

In this paper, we re-examine the theory presented
in~\cite{Graesser:2005ar} using a lattice regulator. This regulator
is non-perturbatively valid and preserves unitarity. We will see that
there is never a ghost when the theory is regulated in this way. As a
consequence, it seems that the conclusions of~\cite{Graesser:2005ar}
are regulator dependent.

\section{Computation}

We begin by describing the theory we will be working with in more detail. The candidate ghost field,  $\phi$, must have a shift symmetry
so it is natural to suppose that it is a Goldstone boson associated with the breaking of some $U(1)$ symmetry. Hence, following~\cite{Graesser:2005ar}, we choose as the bosonic part of the Lagrangian the usual spontaneous symmetry breaking Lagrangian for a complex scalar field $\Phi$,
\begin{equation}
 \cL_b = \partial_\mu \Phi^* \partial^\mu \Phi - \frac{\lambda}{4} \left( \left| \Phi \right|^2 - v^2 \right)^2.
\end{equation}
The Goldstone boson, $\phi$, associated with the spontaneous symmetry breaking is the candidate ghost field.
We couple $\Phi$ to two families of fermions $\psi_i$, $i=1, 2$ of charges $+1$ and $-1$ respectively. We will assume that there are $N$ identical fermions in each family, and that each fermion has the same mass $m$. The fermions are coupled to $\Phi$ by a Yukawa term with coupling $g$. Hence, the total Lagrangian density is
\begin{equation}
\cL = \cL_b + \sum_{j=1}^N \left[ \sum_{i = 1, 2} \left(i \bar \psi_i^{(j)} \gamma^\mu \partial_\mu \psi_i^{(j)} - m \bar \psi_i^{(j)} \psi_i^{(j)} \right) - g \Phi \bar \psi_2^{(j)} \psi_1^{(j)} - g \Phi^* \bar \psi_1^{(j)} \psi_2^{(j)}  \right].
\end{equation}

The low energy effective action for $\Phi$ is obtained by integrating the fermions out.
The effective action can be written
\begin{equation}
\cL_{eff} = \Phi^* G(-\partial^2) \Phi - V(|\Phi|)
\end{equation}
where
\begin{equation}
G(p^2) = p^2 + g^2 N f(p^2).
\end{equation}
The function $f(p^2)$ describes the effects of the quantum corrections to the bosonic kinetic term. If $G(p^2) < 0$ for some range of $p^2$, then the theory can have a ghost. This can only happen if $g^2 N f(p^2)$ is negative and larger than the tree level term $p^2$. Since this signals a breakdown in perturbation theory, we work in the large $N$ limit with $g^2 N$ fixed to maintain control over the calculation.

\begin{figure}
\includegraphics[width=0.5\textwidth]{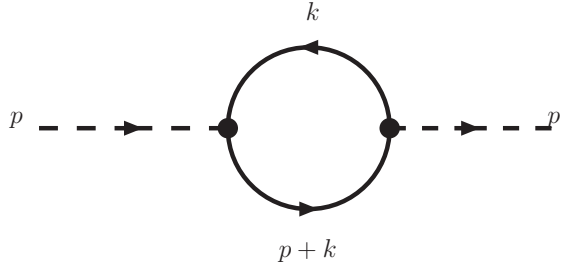}
\caption{The relevant Feynman graph. Dashed lines represent the boson while full lines are the fermions.}
\label{fig:loop}
\end{figure}

Let us now move on to compute $f(p^2)$. To do so, we must evaluate the Feynman graph shown in Figure~\ref{fig:loop}.
After Wick rotating both momenta into Euclidean space, we find
\begin{equation}
f(p^2) = - 4 \int \frac{d^4 k}{(2 \pi)^4} \frac{m^2 - k \cdot (p + k)}{(k^2 +m^2)((p+k)^2 + m^2)}.
\label{eq:unregG}
\end{equation}
This expression is divergent and requires regulation. We choose a lattice
regulator with lattice spacing $a$.  Since we are working in the large
$N$ limit, the phenomenon of fermion doubling~\cite{Montvay:1994cy} will
not pose a problem.  Therefore, we will use naive lattice fermions. The
(Euclidean) fermion propagator is given by~\cite{Montvay:1994cy}
\begin{equation}
G(p) = a \frac{- i \sum_\mu \gamma_\mu \sin (p_\mu a) + m a}{\sum_\mu \sin^2 p_\mu a + m^2 a^2}
\end{equation}
where $\gamma_\mu$ are Euclidean gamma matrices. On this lattice, momentum components lie in the first Brillouin zone, so $ -\pi < p_\mu a < \pi$. The regulated Feynman graph (fig.~\ref{fig:loop}) is
\begin{equation}
f(p^2) = - 4 \int_\mathcal{B} \frac{d^4 k}{(2 \pi)^4}a^2 \frac{m^2 a^2- \sum_\mu \sin a k_\mu \cdot \sin a (p_\mu + k_\mu)}{\Big[\sum_\nu \sin^2 a k_\nu + m^2 a^2\Big]\left[\sum_\rho \sin^2 a(p_\rho+k_\rho) + m^2 a^2\right]}
\label{eq:regulatedG}
\end{equation}
where the integral is over the Brillouin zone $\mathcal B$. Note that as $a \rightarrow 0$, the regulated expression Eq.~\eqref{eq:regulatedG} reduces to the continuum expression Eq.~\eqref{eq:unregG}. 

In~\cite{Graesser:2005ar}, there was a ghost at the origin. Since our goal is to check for potential regulator dependence of this statement, it suffices to extract the order $p^2$ part of $f(p^2)$.
Thus, we expand Eq.~\eqref{eq:regulatedG} in $p_\mu$ and extract the second order term. 
We find
\begin{multline}
f(p^2) \simeq -4 a^2  \sum_\mu p_\mu p_\mu a^2 \int_{\mathcal B} \frac{d^4 k}{(2 \pi)^4} \frac{a^2}{(m^2 a^2 + \sum_\nu \sin^2 k_\nu a)^2}\left[ - \frac{m^2 a^2 - \sum_\nu \sin^2 k_\nu a}{m^2 a^2 + \sum_\nu \sin^2 k_\nu a} \cos 2 k_\mu a  \right. \\
\left. + \frac{1}{2}  \sin^2 k_\mu a 
+ \frac{1}{2} \frac{\sin^2 2k_\mu a}{m^2 a^2 + \sum_\nu \sin^2 k_\nu a}
+ \frac{m^2 a^2 - \sum_\nu \sin^2 k_\nu a}{(m^2 a^2 + \sum_\nu \sin^2 k_\nu a)^2} \sin^2 2 k_\mu a \right] .
\label{eq:order2Exact}
\end{multline}

Now, in~\cite{Graesser:2005ar}, the sign of the kinetic term was altered if the fermion mass was taken to be at least of order of the cutoff. For fermion masses large compared to the cutoff, analytic results were obtained demonstrating the presence of a ghost. In our case, we can obtain an analytic result when $m a \gg 1$. In this limit, the coefficient of $p^2$ induced by the quantum correction is given by
\begin{align}
f(p^2) &= -4  a^2 \left( \frac{1}{m^2 a^2} \right)^2 \int_{\mathcal B} \frac{d^4 k}{(2 \pi)^4} \sum_\mu \left[
\frac{1}{2} p_\mu p_\mu a^2 \sin^2 k_\mu a  - p_\mu p_\mu a^2 \cos 2 k_\mu a \right] \\
&= -4 a^2 \left( \frac{1}{m^2 a^2} \right)^2 \frac{p^2}{4 a^2} . 
\end{align}
Rotating back into Euclidean space, we find
\begin{equation}
G(p^2) \simeq p^2 + g^2 N \left( \frac{1}{m a} \right)^4 p^2.
\end{equation}
Clearly, this quantity never becomes negative, so the sign of the kinetic term does not change in this theory, at least when $ma \gg 1$.

\begin{figure} 
   \centering
   \includegraphics[width=5in]{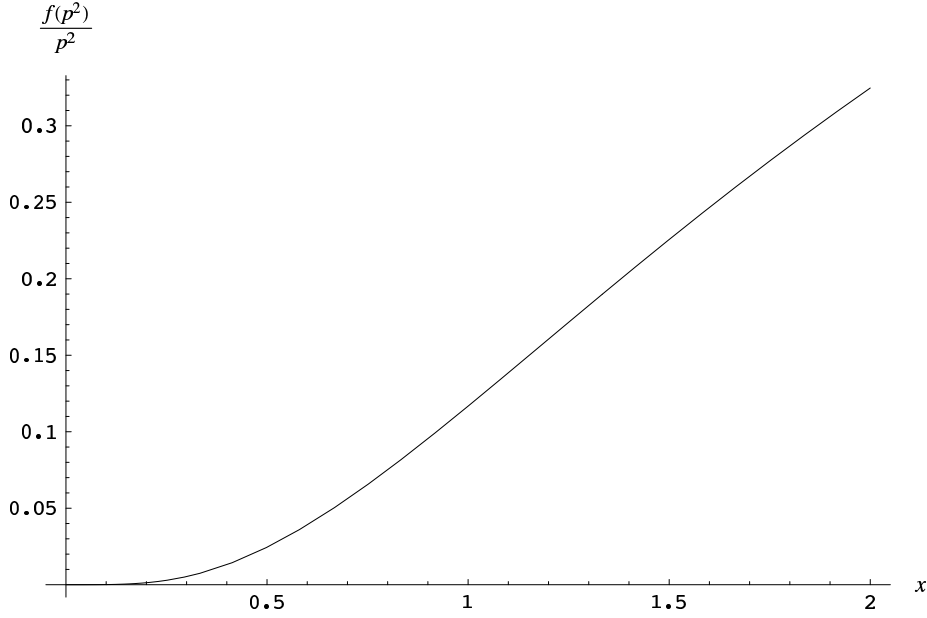} 
   \caption{The quantum correction, $f(p^2)$ as a function of $x=1/ma$. We have shown $f(p^2)/p^2$ for clarity.}
   \label{fig:fofx}
\end{figure}

To check for a sign change away from this limit, we have numerically integrated
Eq.~\eqref{eq:order2Exact} to find the coefficient of $p^2$ induced by quantum corrections, as a function of $x = 1/(ma)$. The result is shown in Figure~\ref{fig:fofx} for $0 \le x \le 2$. Evidently, $f(p^2)/p^2$ is never negative, so there can be no change in the sign. For large $x$, the fermion mass is much smaller than the cutoff so we need not worry about regulator dependence; therefore, we know from the results of~\cite{Graesser:2005ar} that there is no ghost in the region $x>2$. This completes our demonstration that the sign of the kinetic term is always positive if the theory is regulated on a spacetime lattice.

\section{Conclusions}

We have examined the proposed high energy completion of the ghost
condensate~\cite{Graesser:2005ar}. Using a lattice regulator, which
is valid without invoking perturbation theory, and which is unitary,
we have shown that this theory does not have a ghostly low energy
effective action. The effect noted in~\cite{Graesser:2005ar}, which
involved changing the sign of the kinetic term for a scalar $\phi$,
appears to be a regulator dependent phenomenon. Thus, the search for a
UV completion for the ghost condensate must continue.

\acknowledgements

We would like to thank Mark Wise for sharing his insights. We also thank
Michael Graesser and Alejandro Jenkins for helpful conversations and for
proof-reading the manuscript. Finally, we thank \'Ard\'is El\'iasd\'ottir
and Andr\'e Walker-Loud for help with the manuscript.

\end{document}